# Perspective: Organic electronic materials and devices for neuromorphic engineering

Sébastien Pecqueur [a], Dominique Vuillaume [a], Fabien Alibart [a,b,*]

[a]: Institut of Electronic, Microelectronic and Nanotechnology, Centre National de la Recherche Scientifique, University of Lille, Villeneuve d'Ascq, France

[b]: Laboratoire Nanotechnologie et Nanosystèmes, Centre National de la Recherche Scientifique, Sherbrooke, QC, Canada

*: corresponding author, Fabien.alibart@iemn.univ-lille1.fr

Abstract:

Neuromorphic computing and engineering has been the focus of intense research efforts that have been intensified recently by the mutation of Information and Communication Technologies (ICT). In fact, new computing solutions and new hardware platforms are expected to emerge to answer to the new needs and challenges of our societies. In this revolution, lots of candidates' technologies are explored and will require leveraging of the pro and cons. In this perspective paper belonging to the special issue on neuromorphic engineering of *Journal of Applied Physics*, we focus on the current achievements in the field of organic electronics and the potentialities and specificities of this research field. We highlight how unique material features available through organic materials can be used to engineer useful and promising bio-inspired devices and circuits. We also discuss about the opportunities that organic electronic are offering for future research directions in the neuromorphic engineering field.

1- Introduction

While the CMOS technologies are reaching physical limitations in terms of performances, new solutions are expected to emerge in the coming years to sustain the development of information and communication technologies (ICT). This tendency represents a real opportunity to shift and explore new computing approaches that would provide not only better performances, but more adapted solutions to deal with the new needs of our societies. These new requirements should provide solutions to numerous big challenges among which energy limitations constraints, management of the high amount of heterogeneous data generated, massive parallelism and heterogeneity of the communication network and of the interconnected electronic devices (i.e. IoT) that are not coped easily by conventional machines. One of the direction that could potentially answer some of these challenges would be to reproduce concepts and features observed in the brain: this extremely low power computing engine is perfectly adapted to deal with heterogeneous data such as sound, vision, or other sensory-like signals that we record and analyze intensively in our everyday life. Furthermore, its plasticity and ability to adapt and learn make this object extremely resilient to changing environments and operating conditions that would be very appealing for the management of the complex network of devices constituting ICT of tomorrow.

This approach should not only be limited to reverse-engineering of the brain or bio-mimetism (which corresponds to reproducing with the highest precision some biological features observed in biological systems by engineering), nor to map in hardware machine learning algorithms that have proven their efficiency for image classification or other machine learning applications. It could be extended to a broader bio-inspired approach that targets the exploration of features or concepts of interest for computing purposes and finding some material implementation with emerging technologies.

Neuromorphic engineering and computing, a term coined by Carver Mead in the 80s, represents the foundation for this direction.[1] This concept was initially based on the analogy between ions flux across the cell's membrane and the CMOS transistor's transconductance, and has been used for the implementation of the first CMOS neuron. One key aspect of neuromorphic engineering is to deeply rely on the device physics as a computational primitive to build complex computing systems. While neuromorphic engineering has been mostly implemented with standard CMOS technologies (so far leading to bio-inspired sensors and neural circuits),[2,3] the field has recently benefited from the emergence of new materials and devices that have provided new opportunities for neuromorphic engineers. Notably, pushed by the drastic requirements in terms of memory density, resistive memory - or memristive devices - have been envisioned for implementing the synaptic weight connection between neurons.[4] More advanced utilization, closer to biological behavior, capitalized on the memory device physics to realize very efficiently and locally learning rules observed in biology such as, for instance, Spike Timing Dependent Plasticity (STDP).[5-7] More recently, nanoscale devices have been also proposed to implement neuron's building block.[8-10] Note that neuromorphic engineering with emerging technologies is now going beyond single neuromorphic devices and first neuromorphic circuits based on emerging nanoscale devices have been successfully demonstrated.[11,12]

Most of the current approaches share the basic principle of using electronic processes obtained by various devices engineering routes to mimic biological processes. Nevertheless, biological systems and circuits issued from microelectronic are intrinsically very different and it is not clear whether standard electronic platforms will be ideal candidates for neuromorphic implementations. In fact, devices and circuits issued from microelectronic have been optimized for serial data transmission and sequential logic applications where ON/OFF switching ratio, speed and reliability are major constraints while neuromorphic platform relax these requirements (i.e. biological computing engines have intrinsically low signal to noise ratio, slow signal propagation and are fault/variation tolerant). As an interesting alternative to silicon-based technologies, this perspective paper proposes to emphasize on organic materials and devices as potential candidates for neuromorphic engineering. First, these materials and devices are recognized to be very versatile engineering platforms: the soft engineering and bottom-up routes used to synthetize and fabricate them can lead to a very large panel of electronic properties relevant for neuromorphic engineering. Secondly, organic materials and devices can gather both electronic and ionic species as mobile charge carriers. In such systems, electronic and ionic processes can be coupled via numerous basic physical interactions, from electrostatic charge polarization to redox charge transfer. These fundamental properties make organic iono-electronic systems very exciting candidates for implementing bio-inspired concepts and could offer a new toolbox for neuromorphic engineering. We will present here few of the basic material properties that have been used for neuromorphic purposes. These properties will be discussed in the light of selected examples that have been proposed recently for neuromorphic implementation. Finally, we will discuss the very under-looked direction of material and device integration at the circuit and system level offered by organic material processing and will hypothesize on the perspectives that they are offering.

2- Organic Materials and Devices in Neuromorphic Engineering

The aim of neuromorphic engineering at the material level is to find media where information carriers are transported in a way that they may enable similar mechanisms ruling the information transport processes in biological media. From a very general viewpoint, various biological features not available in standard electronics can be highlighted. (i) **Duality**: In the synaptic cleft between two neurons, neurotransmitter molecules and ions are both information carriers: neurotransmitters are a family of chemical carriers confined in the synapses, while ions are charged species delocalized across the neural network via the electrolytic medium and cytoplasm. (ii) **Time Dependency**: Transmission of signals between cells through synapses is largely dynamical. For example, neurotransmitters are responsible for important time dependency of the transmitted signals charging/discharging of the pre-synaptic vesicles and post-synaptic receptors with different kinetically-controlled physicochemical processes. (iii) **Chemical diversity**: The distribution in physicochemical properties of the different anions, cations and molecules makes biological computing' nodes a cross-point of different vectors of information, enabling selective processes necessary to interface their rich environment: for sensing exogenous information or transducing it to the body. In that scope, organic electronic materials have shown over the past ten years great promises in emulating these properties and continue nowadays to bio-inspire us. Following the example of synaptic plasticity, we illustrate how organic materials and devices have been used for neuromorphic engineering and how more could be expected in this direction.

2.1. Small-molecule and allotropic-carbon organic semiconductors materials.

One great advantage of organic semiconductors is the possibility to add new functionalities by integrating different materials via soft processes without destroying their electronic transport properties. We present in this section examples of such material hybridization that have led to interesting neuromorphic applications when various electronic mechanisms are advantageously coupled together.

In 2010, Alibart *et al.* demonstrated how to use charge trapping/detrapping to design an organic Synapstor (synapse transistor) mimicking the dynamic plasticity of a biological synapse (figure 1).[13] This device is also called NOMFET (Nanoparticle Organic Memory Field Effect Transistor) and combines in a single structure both a memory effect (by charge carrier trapping in nanoparticles) and a transistor effect (as the channel conductance is field-effect modulated). This device (which is memristor-like) mimics short-term plasticity (STP),[13] and STDP:[7] two functions at the basis of learning processes (Fig. 1). A compact model was developed,[14] and an associative memory, which can be trained to present a Pavlovian response, was demonstrated.[15] Although the presence of the gold nanoparticles affects the crystallinity of the pentacene semiconductor, the optimized mobility for the functional biomimetic devices can reproductively reach $10^{-1}$ cm$^2$/V$^{-1}$·s$^{-1}$ (higher than the ionic mobility in water).[16] An electrolyte-gated version of this device was developed for biocompatible applications (EGOS: electrolyte-gated organic synapstor).[17] STP with a useful relative amplitude (30-50% of the average DC current) was demonstrated at spike voltages as low as 50 mV, with a dynamic response in the range of tens of ms in aqueous saline solution and cell culture medium (leading to an energy of ca. pJ/spike). These EGOSs working at low voltages (e.g. 50 mV) have performances that open the potentiality to directly interface real neurons. Human neuroblastoma stem cells successfully adhered, proliferated and differentiated into neurons on top of the EGOS as monitored by optical microscopy.[18]

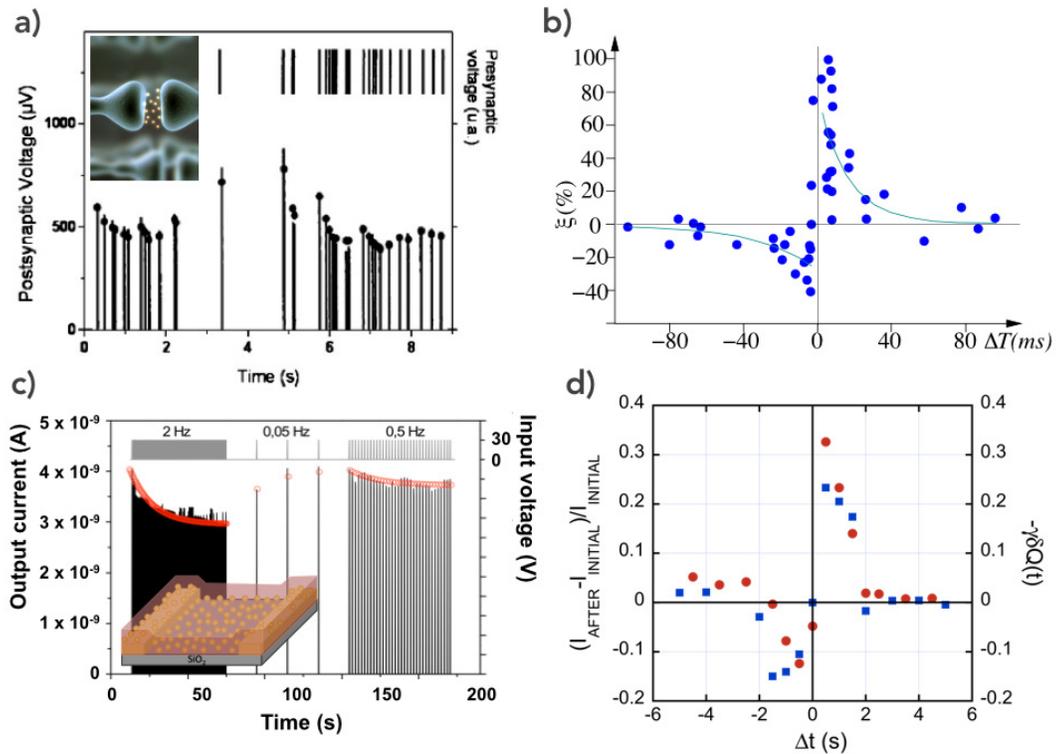

*Figure 1*: *(a) STP (adapted from [19]) and (b) SDTP (adapted from [20]) of a biological synapse and (c-d) corresponding behavior for the NOMFET (adapted from [7,13]).*

Carbon allotropes were also investigated as organic materials for neuromorphic engineering: OG-CNTFET (optically-gated carbon-nanotube field-effect transistor) are interesting since they are programmable (optically and electrically) with multiple memory states.[21-23] OG-CNTFET are made of carbon nanotube FETs (single CNT or random network of CNTs) covered by a photo-conducting polymer (e.g. P3OT, poly(3-

octylthiophene-2,5-diyl)) making them light-sensitive and conferring their non-volatile memory behavior. These carbon-nanotube-based memory elements can be used as artificial synapses and eight OG-CNTFET, combined with conventional electronic neurons, have been trained to perform Boolean logic functions using a supervised learning algorithm.[24] In a crossbar architecture, the OG-CNTFET allowed an efficient individual addressing (reduced crosstalk) by virtue of the development of a "gate protection protocol" exploiting the specific electro-optical behavior of these devices.[25]

Thin films of redox molecules can also be used for neuromorphic devices. Y.P. Lin *et al.* have developed a non-volatile nanoscale organic memristor based on electrografted redox complexes (iron(II) tris–bipyridine complex with diazonium grafting functions) on metal electrodes.[26] These devices are programmable with a wide range of accessible intermediate conductivity states. These authors experimentally demonstrated a simple neural network combining four pairs of organic memristors as synapses (and neurons made of conventional electronics) capable of learning functions.

2.2. Electrolyte-Gated Semiconducting Polymers

Polymers are well established materials in microelectronic fabrication since they can be chemically tuned for their micro/nano-patterning on silicon: they are therefore good candidates for technological hetero-integration on CMOS. Also as organic electronic materials, they offer more flexibility than small molecules to interface solvents for solution processing and liquid electrolyte-gating. The possibility to integrate electrolytes (as ion carriers) on top of organic semiconductors (as hole/electron carriers) without damaging the one or the other offers the possibility to benefit of both charge carrier processes in an all-organic neuromorphic device.

To the best of our knowledge, the first report of an organic device for neuromorphic system was proposed in 2005 by Erokhin *et al.*[27] The device was based on an HCl p-doped polyaniline (PANI) conducting polymer interfaced with a LiClO$_4$/poly(ethylene oxide) (PEO) solid electrolyte.[28] In this device, the kinetically limited process is attributed to a Resistive/Capacitive (RC) internal circuit, correlated to the bulk conductivity of the doped/dedoped-polymer and the accumulation of ions at the PANI/PEO interface.[29,30] The doping of PANI with stronger acids and bulkier counter-anions promotes the repeatability of the device performances while controlling its RC properties.[31] More recently, Demin *et al.*[32] used these devices as the synapse layer of a simple perceptron, which has learned using an error-correction-based algorithm proposed by Rosenblatt in its seminal paper on perceptron,[33] to implement the NAND or NOR logic functions as simple examples of linearly separable tasks. Undoped hole transporting polymers have also been tested such as poly(3-hexylthiophene) P3HT, showing paired-pulse facilitation (PPF),[34] and shows that the formation of water channels in the solid electrolyte plays an important role in the functioning.[35]

2.3. Iono-electronic polymers: mix electronic and ionic conduction in organic materials

A more intimate coupling between ions and electrons can be obtained when their interaction is not limited at the interface between two materials but could be realize in the bulk. This option has been advantageously deployed to transduce ionic into electronic signals (and reciprocally) in a variety of organic electronic materials.

Iono-electronic polymer can be intimate blends of charged polymers, such as the well-known poly(3,4-ethylenedioxythiophene):poly(styrenesulfonate) (PEDOT:PSS). The semiconducting PEDOT is a low oxidation potential polymer which thermodynamically undergo in its oxidized state in moist air.[36] In PEDOT:PSS, the PSS-negative charges are compensated by a positive charge (i.e. a hole) on the PEDOT aromatic molecule to respect electro-neutrality. The negative charges on PSS$^-$ are fixed while the positive charge on the PEDOT$^+$ can easily be delocalized and contribute to electronic transport. In addition, if some external mobile ions can penetrate into the bulk of the PEDOT:PSS, the local electronic conductivity will be tuned based on the same electro-neutrality principle (i.e. adding one monovalent positive ion balances one hole extracted from the PEDOT polymer chains) . The abilities of these materials and devices to gradually change their electrical conductivity upon ion/electron exchanges made them promising materials for brain-like circuitry that is in nature an "iontronics" system. Li *et al.* used PEDOT:PSS in a rectifying memristor structure to demonstrate STP, long-term plasticity (LTP), STDP and spike-rate-dependent plasticity (SRDP).[37] It has also been used with liquid-electrolyte gating, showing time-dependent paired-pulse depression (PPD).[38] By substituting the PSS to poly(tetrahydrofuran) (PTHF), the memory becomes less volatile,[39] and promotes LTP,[40] with PPD about ten times slower than PEDOT:PSS systems.[38] Based on PEDOT:PSS, another device structure named ENODe (electrochemical neuromorphic organic device) was recently proposed as a low-voltage organic synapse.[41] The device structure features two PEDOT:PSS electrodes (one partly reduced by a poly(ethylenimine) treatment separated by an electrolyte. Upon application of pulse voltage on one of the PEDOT:PSS electrode (used as pre-synaptic input), cation exchange through the electrolyte modifies the conductivity of the PEDOT:PSS/PEI film used as the post-synaptic output. Working at a low voltage (around 1mV) and a low energy (10 pJ), this device showed non-volatility, long-term potentiation and depression with 500 discrete states within the operating range.

2.4. Perspectives

On the first aspect about the information **_carriers' duality_**, electrolyte-interfacing organic semiconductor systems can mimic biological synapses with electrons/holes emulating the ions and ions emulating the neurotransmitters (Figure 2a). The slow dynamics associated to ion/electron interaction are well adapted to reproduce volatile memory effects. This represents an advantage over other standard electronic technologies, in particular RRAM or OxRAM systems that have been optimized for non-volatile memory applications.[42] While memristive devices have been mostly considered for their analog programmability potential (i.e. implementing advantageously synaptic weight), such volatile mechanisms are only obtained in diffusive memristors[43,44] with only little possibilities to be adjusted (i.e. unstability in nanoscale filaments). Advantageously, volatility level appears to be largely tunable in organic system. In addition, studies performed with solid electrolytes such as PEO shows that these platforms can be downscaled for high-density development,[45,46] while studies performed with water as an electrolyte shows their potential application to interface biology.[47-49]

The **_time-dependency_** of signal transmission and propagation observed in biology can also be advantageously reproduced with organic systems. For instance, ionic and electronic properties such as charge mobility in electrolyte or metal, respectively, are strongly different and result in very different performances for signal propagation. Consequently, implementing at the device level bio-realistic time constant observed in biology (such as membrane time constant in neural cells, or neurotransmitter dynamics at the synaptic cleft) requires large capacitances in the purely electronic medium and are a severe limitation in terms of circuit design. In particular, the interplay between ionic transient currents and electronic steady-states allows the tuning of the signal propagation dynamics,[50] which influences all the plasticity-related elementary mechanisms of the artificial synapse: the interfacing of the electrolyte with the semiconductor is therefore a key parameter. Iono-electronic polymers have the advantage to be hydrophilic compared to neutral polymers and allow their swelling in the presence of water,[51] promoting the ion charge/discharge of the bulk of the material rather than the top surface.[52] The bulk capacitance of these electrolyte-interfacing materials is non-ideal and relates to constant-phase-elements impedances of porous systems, which can be emulated with infinite numbers of series/parallel RC elements (Figure 2b).[53,54] Analogously, these non-ideal impedances are also ruling the diffusion-control existing at the cellular membrane.[55-57] Developing further organic semiconductor promoting the ion transport at nanoscale in the bulk of the electrical material is attractive to emulate the non-ideality of the impedance ruling the whole time-dependency in the electrochemical signal response. Other ionomers are of recent interest in organic electrochemical transistors such as conjugated polyelectrolytes (either self-doped or intrinsic)[58-60]. Also several works are currently promoting the conception of neutral semiconducting polymer with hydrophilic properties, by the introduction of glycol chains.[61,62]

Finally, at the **chemical diversity** level, organic electronics could offer a well-adapted platform to couple various functionalities at the device level. Coupling neuromorphic sensing, transduction and computing at the hardware level is still not extensively investigated, although biomimetic sensing/transduction platforms are in need for neuromorphic data analysis of pattern-recognition based applications.[63,64] Especially for input/output layers, sensing/transduction of exogenous information (light, mechanical deformation, chemical) to/from ionic action potentials requires specific materials for coupling the neural information transport properties to the desired sensed/transduced physical information. Organic semiconductor materials can show fluorescence and phosphorescence,[65] promoting a direct coupling between the material electrical properties and a specific wavelength range of emitted light (as organic light-emitted diodes)[66,67], or absorbed one (as organic photodiodes).[68-70] The photo-transduction associated to the modification of the rectification ratio of photodiodes by light can modulate the material electrical conduction under reversed bias, mimicking retinal photoreceptor cells. Organic semiconductors are also mechanically stable[71], and semiconducting iono-electronic polymers systems can also be electroactive and transduce by electrical stimuli as low as 2 V to morphological changes.[72] Reversibly, they can also sense mechanical deformations (as cutaneous nerves or cochlea's hair cells), modifying the electrical properties of electronic skins sensors.[73,74] Finally, electrical properties of organic semiconductors can also be molecularly modulated, gating the electronic state of the channel from non-conducting (inhibition) to conducting (excitation). Molecular-gating of the organic semiconductor conductivity can be done either directly in the material by doping with molecularly-specific strong electron acceptors (p-type) or donors (n-type),[75,76] or indirectly by performances modification of an external gate with specific electroactive agents contained in the electrolyte.[77,78] The integration of such molecular agents in the material (either the semiconductor or the electrolyte), modulating the conductivity accordingly to their chemistry, provides a versatile mean to incorporate both performance variability and selectivity necessary to integrate multiple processes at the network level using a single electronic technology.

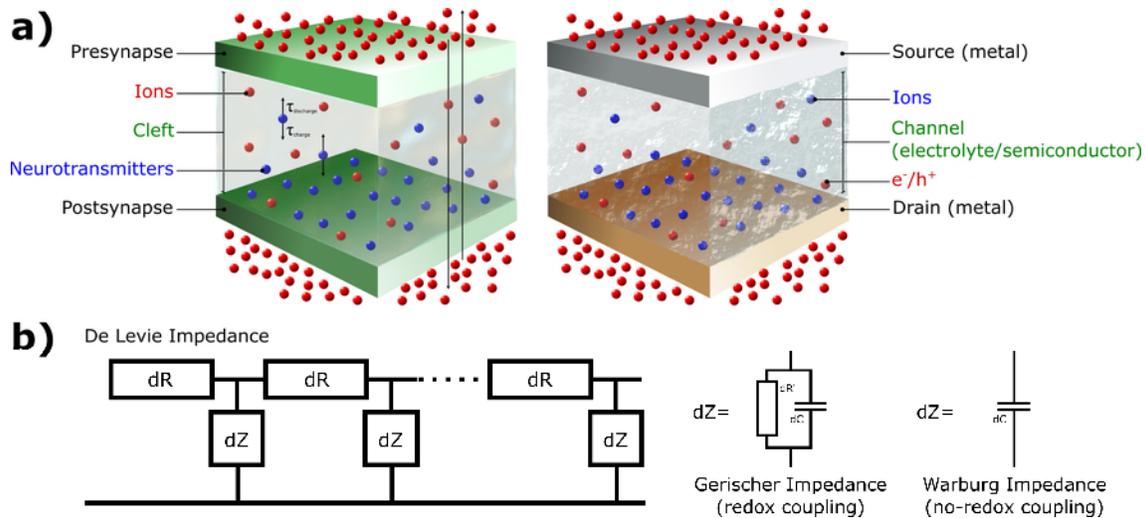

*Figure 2: (a) Analogy between a biological synapse and organic-electrochemical synapse. (b) De Levie infinite transmission line impedance model for porous electrode with redox (Gerischer Impedance) or without redox (Warburg Impedance) reaction.*

3- Circuits and systems based on organic processes

3.1- General considerations and recent achievements

One of the important challenges that neuromorphic engineering is facing is to bring elementary devices (or basic building blocks) at the circuit and system level. Some important aspects of the problem are: (i) **To reproduce the high parallelism of neural network in hardware**. For example, conventional memory structures are mostly accessed sequentially when one would deeply benefit from a parallel and distributed memory architecture to emulate synaptic operations between cells. Potential solutions to this challenge have been considered based on the idea of passive crossbar arrays with resistive memories. But, scaling of this approach seems limited due to physical limitation in terms of energy cost and engineering limitations when one tries to move charges over long and passive metallic wires[79]. (ii) **To reproduce the combination of both local and global effects observed in biology**. Some processes are highly local such as synaptic plasticity processes (i.e. Spike Timing Dependent Plasticity, for example) and would benefit from embedded plastic features as close as possible to the memory point. Some others such as reinforcement learning (i.e. associated to processes such as dopamine delivery in the brain) or homeostasis are global effects affecting large populations of cells and would benefit from a more global circuitry such as a central processing unit. This later aspect is somehow in opposition with the idea of distributed computing units and locality. (iii) ***To reproduce the complex wiring between populations of cells in hardware***. In fact, if the issue of fan-in / fan-out could be solved, one remaining question is to implement in hardware the ability of cells to transmit information on different length scale. For instance, if one neuron projects on 10000 neurons on average, the target population of this projection varies deeply from cell to cell and involves both "hard-wiring" (i.e. neural topology that pre-exist before learning) and "plastic-wiring" (i.e. wiring between cells resulting from different learning experiences). Since this complex wiring is in principle not known before learning, the conventional top-down technological approach is to oversized the interconnection between cells (i.e. allow for the highest degree of interconnection between cells and restrict it after learning) or to make arbitrary decision on a reasonable degree of interconnection and a pre-defined population of cells that will be allowed to communicate between each other. The first aspect is somehow illustrated by the concept of passive crossbar arrays and the second by the most advanced neuromorphic implementations such as True North[80] or Spinnaker[81] where Address Event Representation (AER) is used to emulate parallel communication between populations.

3.2. Organic materials for dendritic engineering

Passive crossbar arrays of resistive memory cross-point are an elegant solution to reproduce the high level of parallelism between cells[4]. It corresponds to implementing with metallic wires the axono-dendritic tree of biological neural cells and to implement the synaptic weight with memory cross-point. In this approach, one challenge is to minimize wires' resistance in order to not hide the memory elements' resistance itself.[79] Nevertheless, this approach neglects some important aspects at work in biological networks such as dendritic computing.[82] In fact, transmission speed along axons and dendrites, localization of the synaptic cleft along the dendritic arbor and their respective timing are used as important computing ingredients by biological cells. These features can hardly be implemented with purely electronic conductors or at the expense of heavy overhead circuitry to implement the bio-realistic time constant for example. Few interesting propositions have been recently published in the field of organic electronics that could offer new perspectives for this particular feature. Xu *et al.* proposed to implement the interconnecting wires (figure 3a) with conducting organic materials.[83] This approach is of interest for the ease of implementation of the conductive wires by inkjet printing but also lead to interesting dendritic properties. So far, the authors demonstrated PPF and STDP on the organic nanowire/synapse system. Note that the material system is an organic/electrolyte device in nature and that the combination of both ionic and electronic dynamics results in bio-realistic temporal features at the device level. An interesting perspective would be to extend the analogy with dendritic processes at work in biological cells.

Along the idea of dendritic computing features, other works have reported the possibility to use organic materials for implementing dendritic connections. Work from Malliaras *et al.* proposed to implement dendrites with PEDOT:PSS materials.[84] This iono-electronic polymer was used to demonstrate orientation selectivity along a long PEDOT line with multiple gates localized along the wire in an electrolytic environment (figure 3c). Interestingly, time lag effect was demonstrated to depend on the gate location and a direct analogy with temporal integration along the dendrites can be seen in this experiment. Temporal features were the result of both the ionic dynamics in the electrolyte and the electronic properties of the PEDOT:PSS material.

A third material system based on Indium Zinc Oxide (IZO) transistor electrolitically gated with a chitosan membrane demonstrates dendritic features (figure 3b).[85] In this work the chitosan layer is used as a proton conductor material and can be thought as equivalent to an artificial dendrite. The organic ionic conductor (chitosan) is then used to implement temporal features observed in biological dendrites with multiple gates implementing the pre-neuron input.

Finally, toward the idea of mimicking biology, Yang *et al.* reported on the realization of an ionic cable based on polyacrylamide hydrogel (figure 3d).[86] This purely ionic system based on ionic conductors separated by an insulator was used to implement an ionic propagation line. This system is a direct analogy to the neuron's membrane and to the way spikes propagate along the axono-dendritic tree.

We believe that organic materials and their ubiquitous ionic / electronic properties represent a real opportunity to implement such dendritic processes and that more research efforts in this direction could lead to very appealing technological solutions for neuromorphic engineers.

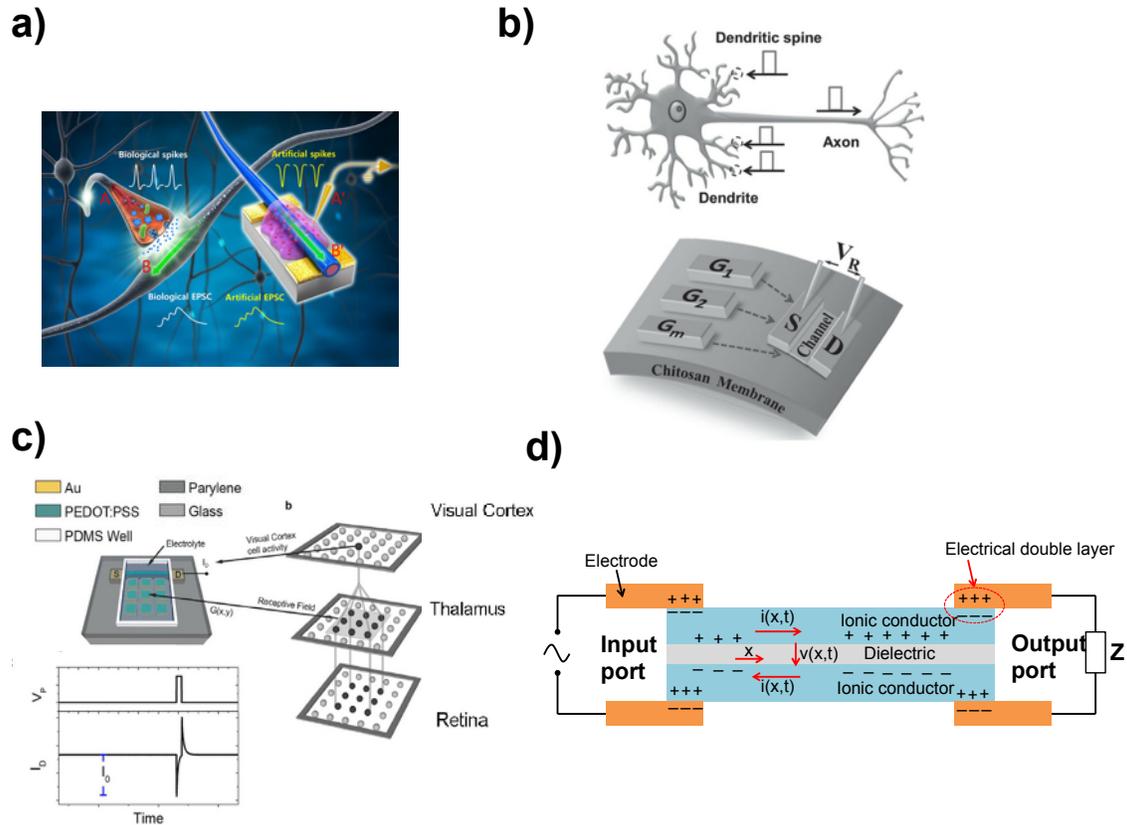

*Figure 3: (a) Organic dendrites and synapses realized by inkjet printing of organic core-sheath nanowire (from [83]). (b) Emulation of dendritic propagation of signals with IZO transistor and chitosan ionic membrane (adapted from [85]). (c) Orientation selectivity with dendritic-like organic electrochemical transistors (adapted from [84]). (d) Concept of ionic cable (adapted from [86]).*

3.3. From local to global processes: mix ionic/electronic conductors with global electrolyte

As mention previously, learning in biology can result from different spatial scales. From purely local event induced by chemical processes happening at the synaptic cleft, to extended contribution such as the tri-partite synapse that involve the contribution of the glial cells into synaptic potentiation (note that glial cells can extend along multiple neurons and provide subsequently an additional interconnection between these cells), to reinforcement learning where the overall synaptic learning can be influenced by a global marker such as dopamine concentration. These mechanisms are possible thanks to the multiple carriers of information that are used in biological engines (various ions and chemicals) and that are hardly reproduce with single data carriers in standard electronics. Gkoupidenis *et al.* published an interesting solution toward this goal recently.[87] The proposed system was composed of a passive matrix of OECTs interconnected by metallic wires implementing the "electronic" part of the system and of a global electrolyte used to gate the OECT devices implementing the "ionic" part of the system (figure 4a and 4b). Local synaptic effects can potentially be implemented on single OECT (see previous sections) while global effects are implemented via the electrolyte and the ionic carriers of information. Integrating a grid of 4x4 OECTs gated by the same electrolyte, these authors demonstrated that they can globally control the behavior of the OECTs in the array, a behavior that resembles homeoplasticity phenomena of the neural environment. They also show that OECTs can communicate though the common electrolyte, i.e. an input signal sent to one OECT is also detected by a neighboring one in the network.

Along this line, we propose recently a concept of spatial reservoir with OECT sensors.[88] In this system (figure 4c-e), the connectivity through the electrolyte was used to transmit the input signal from a global gate to a network of 12 OECTs. Thanks to the electropolymerization of the p-type iono-electronic polymer, each individual sensing element presented a many-fold variability in time response, charge/discharge-symmetry, channel conductance and transconductance, affecting the output image of the gate input signal. We demonstrated in this work that non-trivial binary classification tasks can be realized out of this system: the system has been able to discriminate two different frequency-modulated voltage pulses patterns injected from the gate, with error rates controlled by the number of training vectors, the number of OECTs and their variabilities. Showing that the more the better in terms of number of OECTs and training vectors, this study also unravels that an appropriate set of a restricted number of OECTs can systematically lead to an optimal recognition or not, depending on the individual device properties (the relevant properties responsible for pattern recognition performance of the network have not been quantitatively identified yet). Considering the relationship between electrolyte nature and the gate field-effect efficiency[75,89], more is expected in a study where a fixed and periodical pattern would be injected from the gate and the electrolyte medium would be modified as an input: more complex interaction between the electrolyte and OECT such as sensitivity to various ionic species, sensitivity to concentration could be added into the overall system.

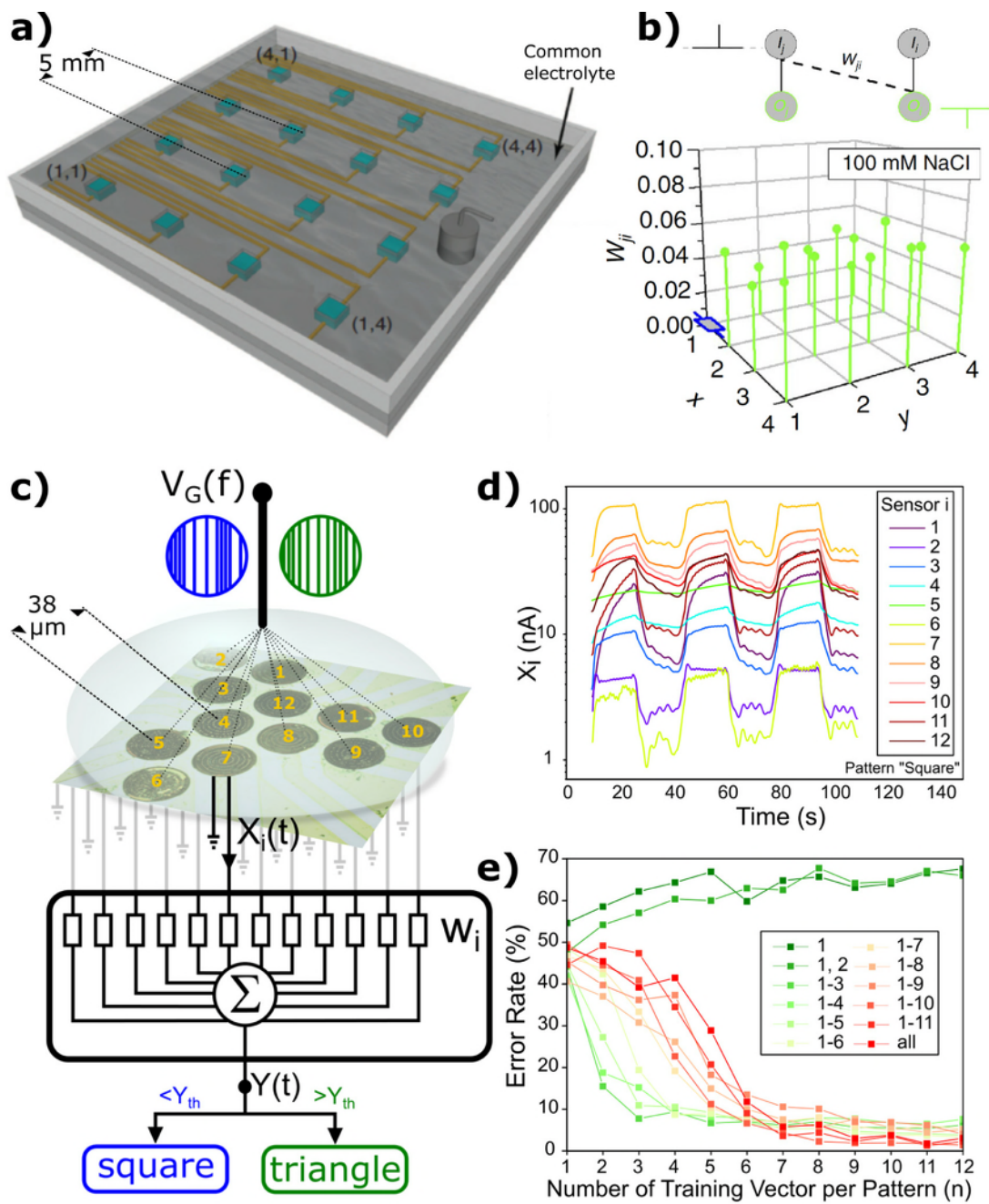

*Figure 4*: *(a-b) Global gating through the electrolyte of organic electrochemical transistors showing both local and global tenability of the devices (adapted from [87]). (c-d) Concept of spatial reservoir implemented with electropolymerized OECTs for pattern classification (adapted from [88])*

3.4. From synaptic learning to network wiring.

As pointed out by J. Hawkins at IEDM 2015,[90] synaptic weighting with analog weight such as resistive memory is in essence different from the concept of wiring observed in biology. The former corresponds to continuous evolution of the synaptic weight between two values (the min and max conductance of the memory element) while the later corresponds to a continuous evolution of an "undisplayed" synaptic state. In other words, the potential synapse that will result from the growth of a dendritic path does not contribute to computing until the synaptic connection is established. Nevertheless, important learning and computing events can be attributed to this dendritic growth. One could expect that artificially reproducing the dendritic wiring through fixe network such as parallel crossbar topologies will miss important aspect of biological computing, notably in terms of energy consumption since all weights in the network will contribute to power consumption (even if there conductance is small) while undisplayed synaptic element not. In addition, aging consideration should consider that if fix topology network can be used to compensate to device aging (i.e. synaptic connections that become inefficient are replace by new ones) to some extend (until enough alternative pathway are available), this approach does not solve completely the important ability of biology to regenerate new connections.

3.5. Perspectives for neuromorphic

Organic electronic materials offer also biomimetic perspectives at the network level by their fabrication processes (top-down and bottom-up), which could emulate the morphological evolution of the neural network while preserving the functionality of the elementary organic electronic elements.

On the nucleation, self-assembling of conducting polymers micro/nano-object could rapidly access to the systematic patterning of high density networks of functional conducting polymer units. Prior assembling, conducting polymers have been can be formulated into nanoparticles and micro-spheres of different dimensions by many different template-free versatile techniques such as droplet microfluidics (50 µm in diameter),[91] delayed precipitation (diameter between 1 and 10 µm),[92] sono-electrochemical polymerization (diameter between 1 and 4 µm),[93] and by dispersion polymerization (diameter between 0.2 and 2 µm).[94,95] The combination of these top-down approaches offers large perspectives of active-material co-integrations, anisotropic assembling variabilities and three-dimensional interconnection for high density networking.

On the growth and interconnection, electropolymerization is a voltage-guided synthetic process which allows the nucleation and growing of conducting polymer objects, potentially providing an intimate coupling between device operation and network modification. As an example, Inagi *et al.* have demonstrated the possibility to synthesize dendritic PEDOT fibers using bipolar electrochemistry[96-98]. In a polarized AC electric field, the conducting polymer fibers bridge neighboring gold wires and propagates from wire to wire along this field. The voltage-guided propagation of the conducting polymer growth promotes the arborescence of a conducting network, while voltage also promotes the information conduction through the network. Therefore, conducting polymer technologies offer the perspective to couple both the physical evolution of the network with data processing. Combining the different soft-techniques of town-down assembling of conducting polymers and bottom-up electropolymerized interconnection could provide an elegant technique to realize such functional bioinspired neural network,[99] but also to have a coupling between network morphological change and operation, necessary to mimic long-term memory at the network level.

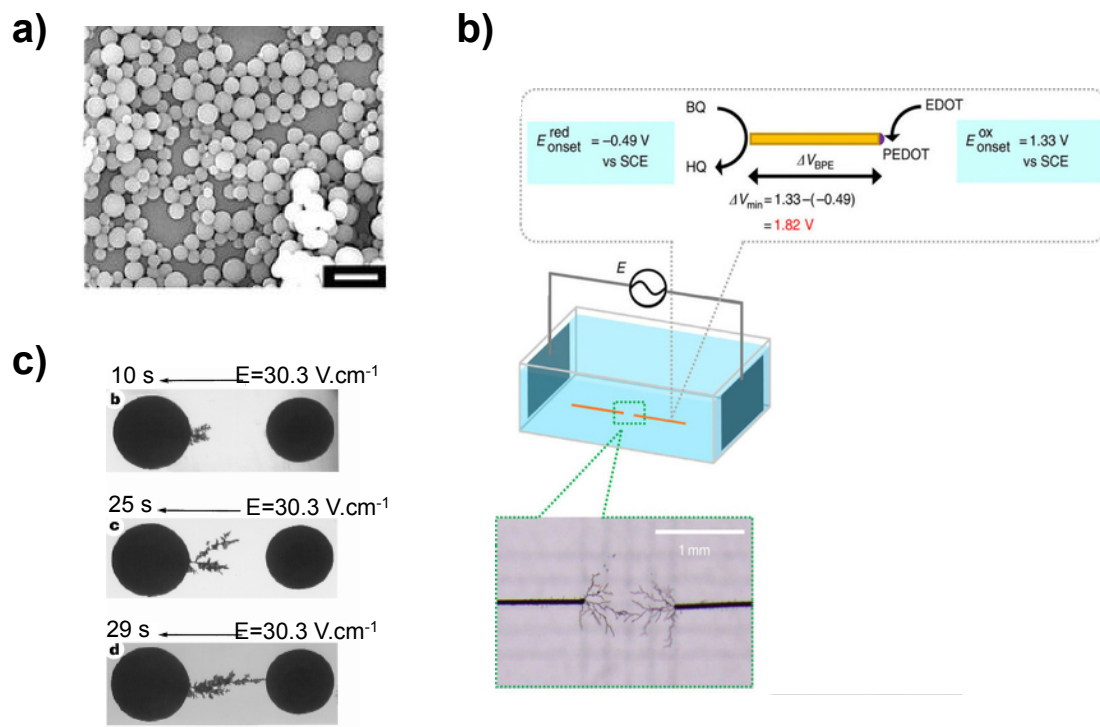

*Figure 5: (a) Assembling of conducting polymer microspheres (adapted from [92]), scale bar is 5um. (b) Dendritic PEDOT:PSS interconnection (adapted from [97]). (c) Electrical interconnection of microspheres (adapted from [99]).*

4- Conclusion

In summary, the recent results reviewed in this paper have demonstrated that organic materials and devices are prone for the implementation of neuromorphic functions, not only at the single device level, but also at the circuit level. Some perspectives are discussed showing the possibility to improve these systems towards an increased complexity in terms of functions and circuit connectivity. From a very general viewpoint, table 1 points out the intrinsic properties of organic iono-electronic materials and inorganic silicon-based technologies. At the light of this table, the different perspectives from materials, devices, to system level for neuromorphic engineering can be evaluated. Since neuromorphic engineering is still an emerging computing paradigm, the question of the most appropriate technology is still completely open. This last statement should be of course balanced by the very important criterion of technology maturity that will enable neuromorphic technologies to be transferred to the industrial level.

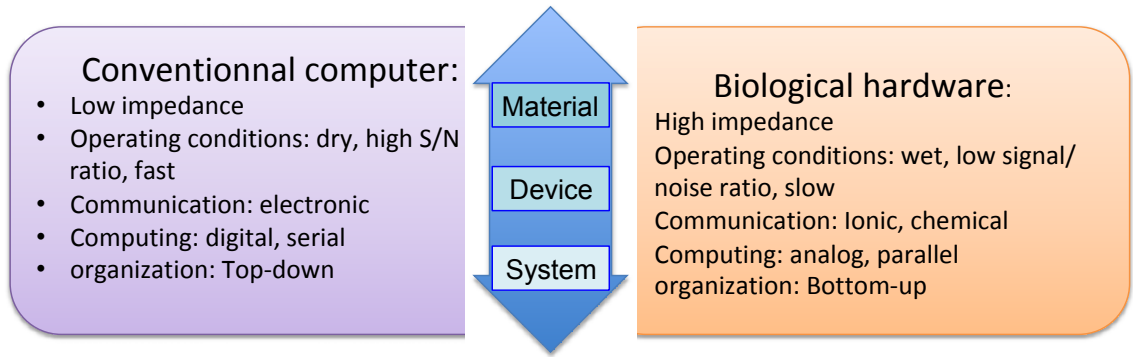

|  | CMOS technology | Organic ion/electronic technology | Biology |
|---|---|---|---|
| Operational environment | Dry (sealed packaged) | Air / water stable | Wet |
| Information carrier | electrons / holes | electrons / holes / ions / chemicals | ions / chemicals |
| Mobility ($cm^2/(V.s)$) | $1-10^3$ (silicon) | $10^{-6}$ to $10$ (electronic) $10^{-4}$ to $10^{-3}$ (ionic) | $10^{-3}$ (ionic) |
| Capacity ($\mu F/cm^2$) | 1 (high-k oxide with thickness of few nm) note: Capa. decreases with thickness | 500 (130 nm PEDOT film with volumic capacitance) note: Capa. increases with thickness | 1 (cell's membrane) |
| Fabrication | top-down lithography | top-down lithography bottom-up self assembling, eletropolymerization | Bottom-up |

Table 1: Comparison of the main intrinsic properties for both organic and inorganic technologies. These properties could be thought in the light of the main characteristics of biological computing systems.

**Acknowledgements.**

Parts of the works reviewed here have been financially supported by the European Union projects NABAB (FP7-FET-FP7-216777), SYMONE (FP7-FET-318597), I-ONE (FP7-NMP-280772), RECORD-IT (H2020-FET-664786) and the French Agency ANR project SYNAPTOR (12 BS03 010 01).

References:

1. C. Mead, *Proceedings of the IEEE* (1990), p. 1629.

2. C. Bartolozzi, R. Benosman, K. Boahen, G. Cauwenberghs, T. Delbrück, G. Indiveri, S.-C. Liu, S. Furber, N. Imam, B. Linares-Barranco, T. Serrano-Gotarredona, K. Meier, C. Posch and M. Valle, *Neuromorphic systems.* (Wiley Encyclopedia of Electrical and Electronics Engineering, 2016).

3. G. Indiveri and S.-C. Liu, *Proceedings of the IEEE* (2018), p. 1379-.

4. D. B. Strukov, Nature **476**, 403 (2011).

5. S. H. Jo, T. Chang, I. Ebong, B. B. Bhadviya, P. Mazumder and W. Lu, Nano Lett. **10**, 1297 (2010).

6. D. Kuzum, R. G. Jeyasingh, B. Lee and H. S. P. Wong, Nano Lett. **12**, 2179 (2011).

7. F. Alibart, S. Pleutin, O. Bichler, C. Gamrat, T. Serrano-Gotarredona, B. Linares-Barranco and D. Vuillaume, Adv. Funct. Mater. **22**, 609 (2012).

8. J. Grollier, D. Querlioz and M. D. Stiles, *Proceedings of the IEEE* (2016), p. 2024.

9. J. Torrejon, M. Riou, F. A. Araujo, S. Tsunegi, G. Khalsa, D. Querlioz, P. Bortolotti, V. Cros, K. Yakushiji, A. Fukushima, H. Kubota, S. Yuasa, M. D. Stiles and J. Grollier, Nature **547**, 428 (2017).

10. P. Stoliar, J. Tranchant, B. Corraze, E. Janod, M.-P. Besland, F. Tesler, M. Rozenberg and L. Cario, Adv. Funct. Mater. **27**, 164740 (2017).

11. M. Prezioso, F. Merrikh-Bayat, B. D. Hoskins, G. C. Adam, K. K. Likharev and D. B. Strukov, Nature **521**, 61 (2015).

12. G. W. Burr, R. M. Shelby, S. Sidler, C. Di Nolfo, J. Jang, I. Boybat, R. S. Shenoy; P. Narayanan, K. Viwani, E. U. Giacometti, B. N. Kürdi and H. Hwang, IEEE Trans. Electron Devices **62**, 3498 (2015).

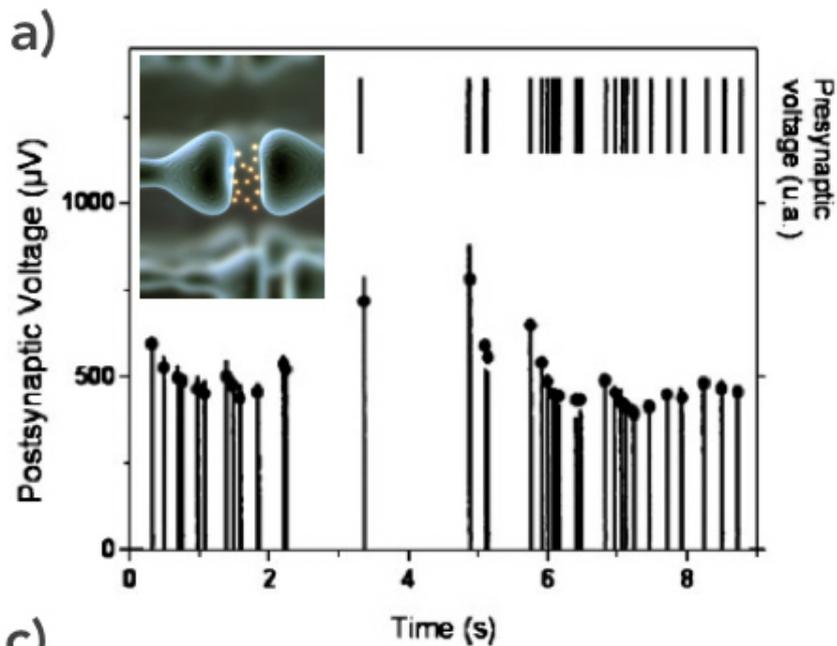
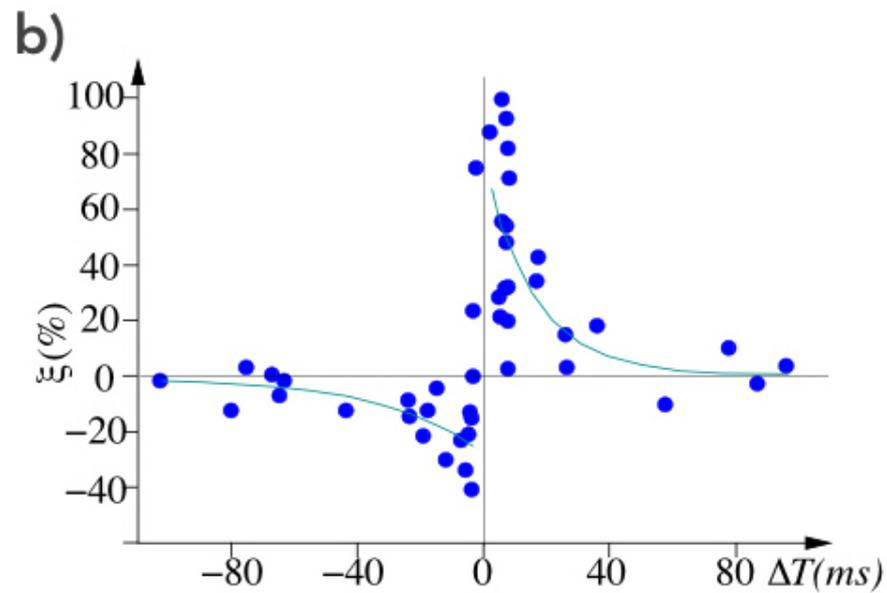
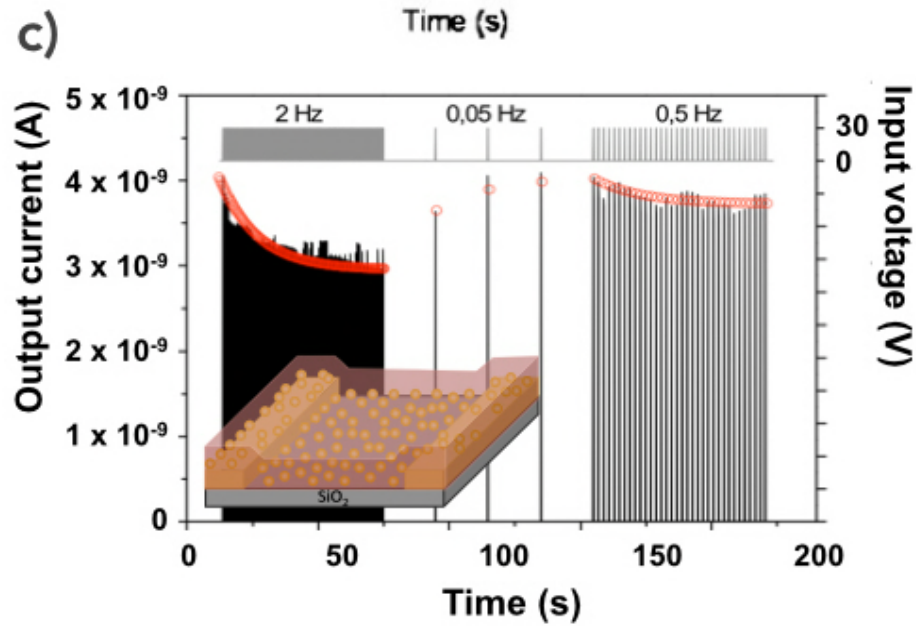
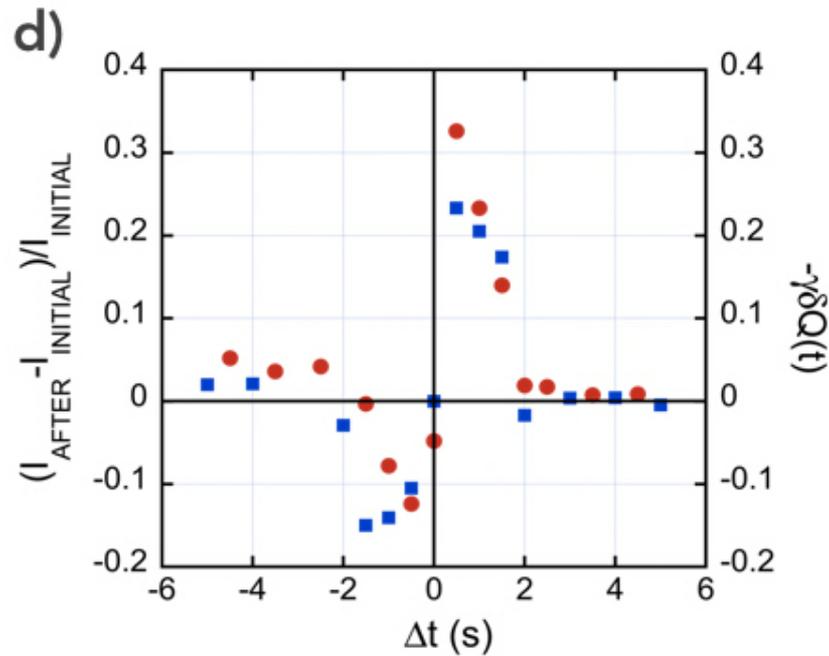

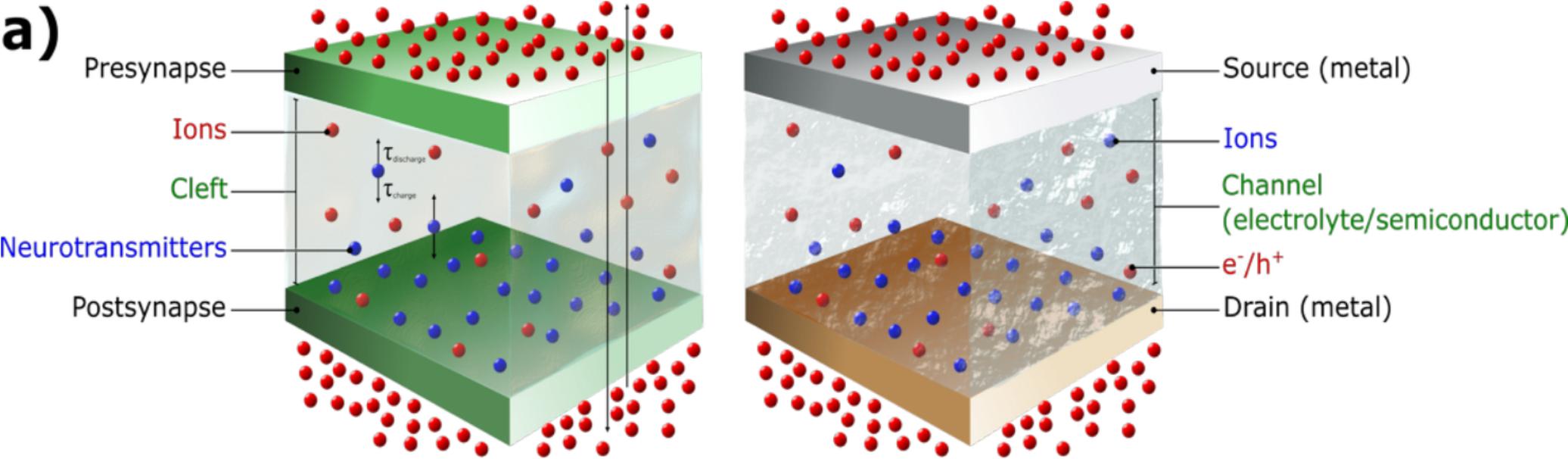

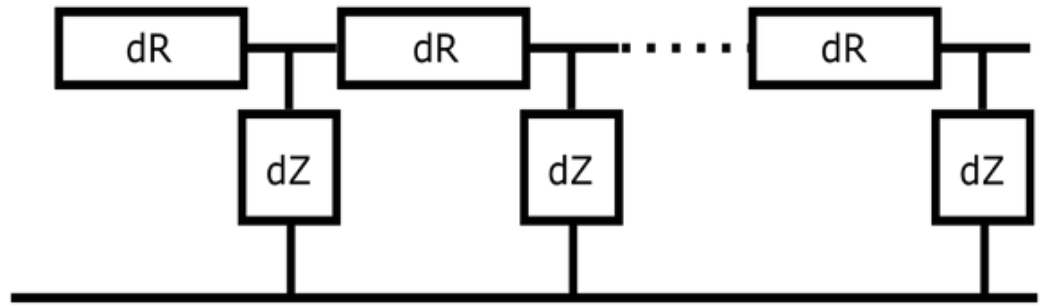
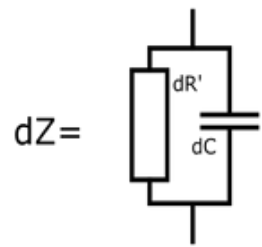
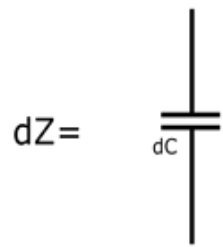

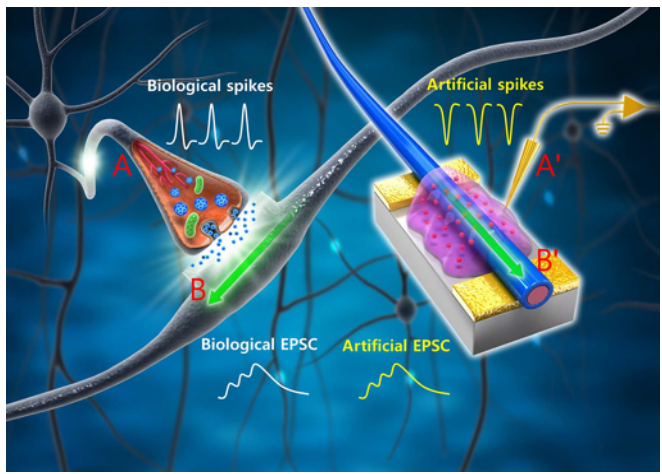
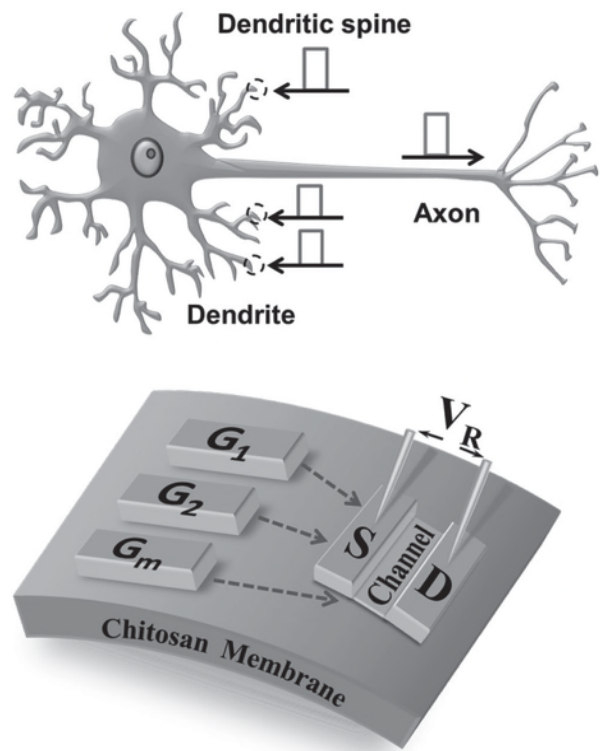
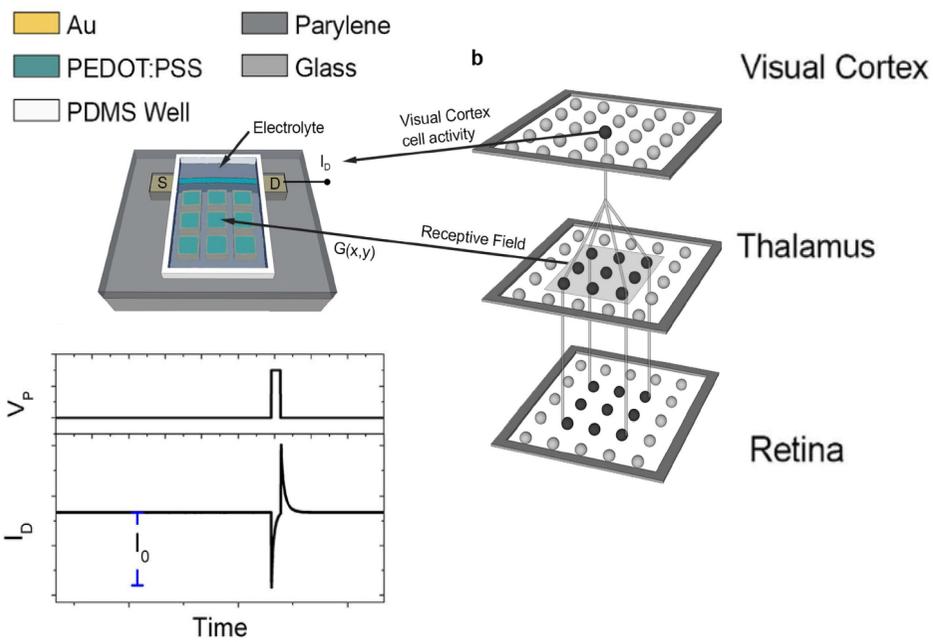
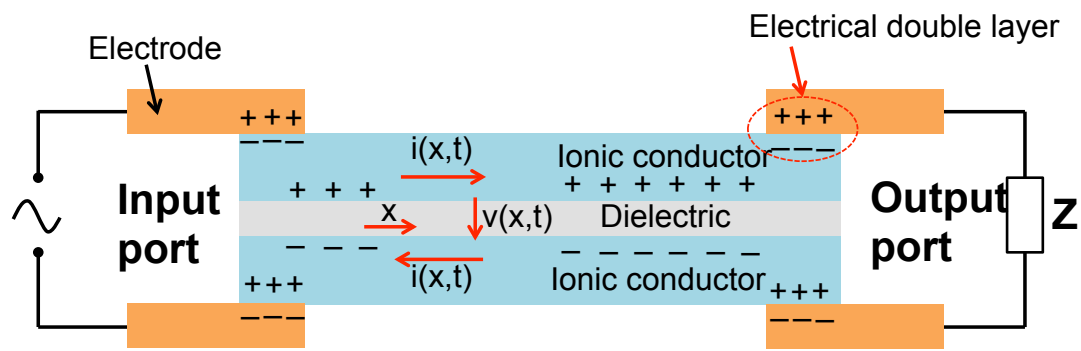

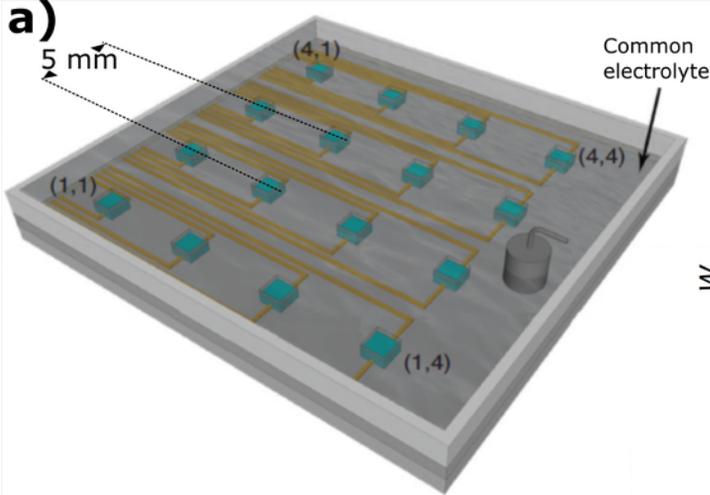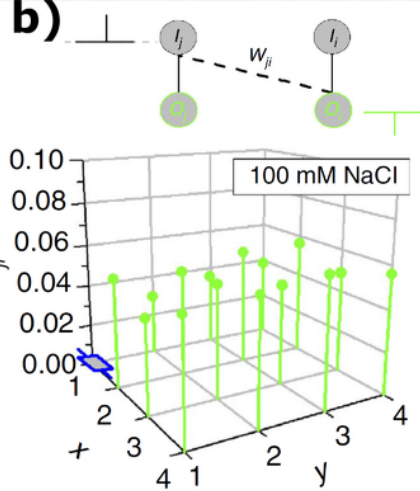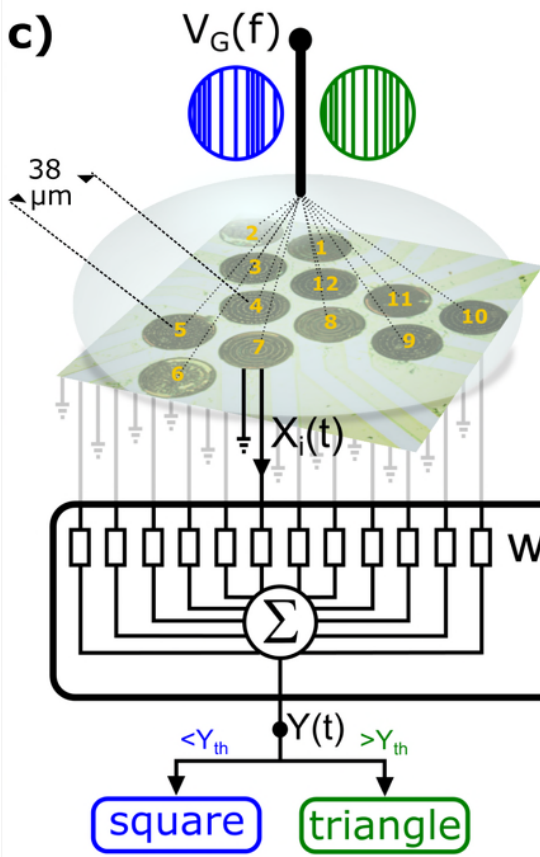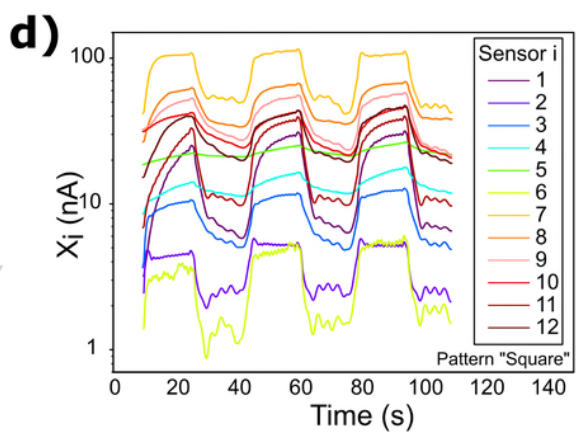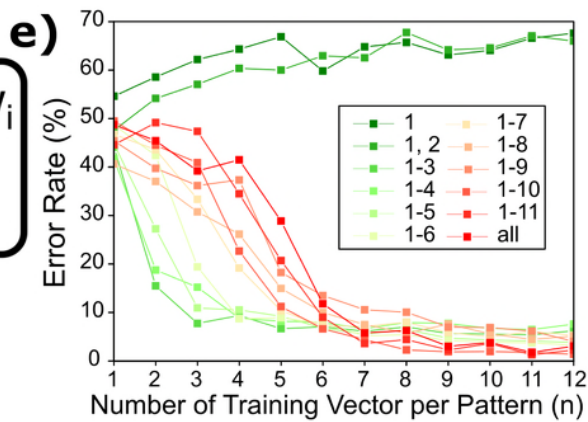

**a)**

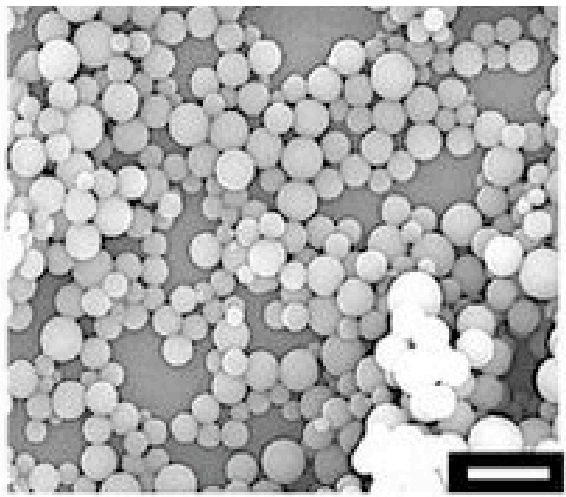

**b)**

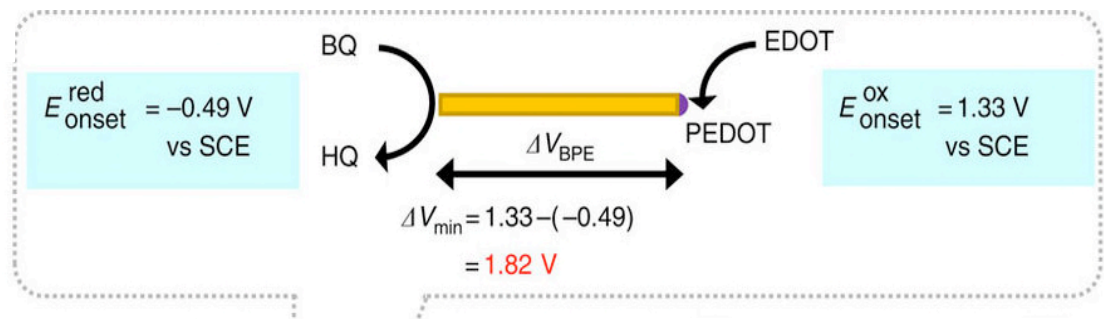

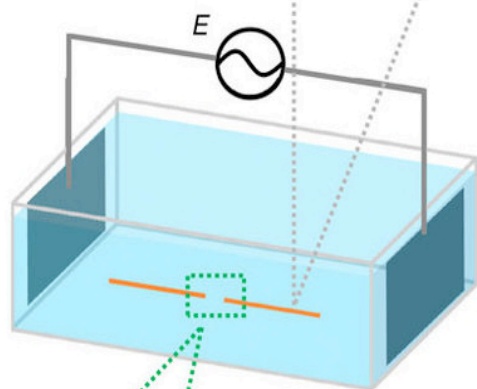

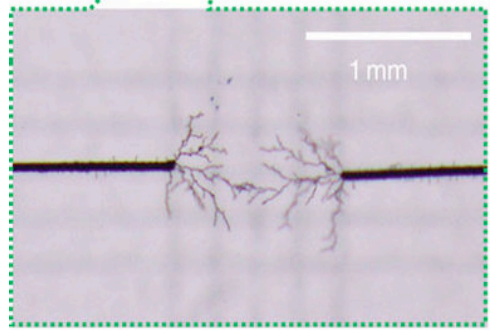

**c)**

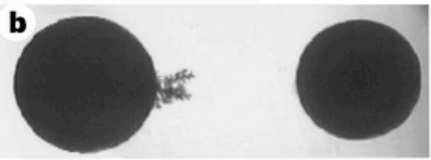

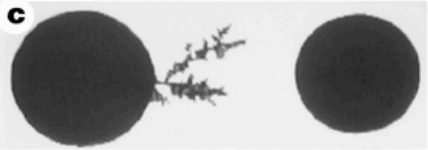

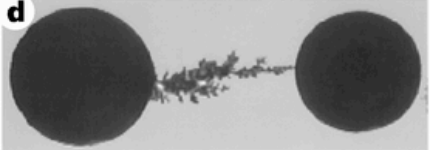

**Conventionnal computer:**
- Low impedance
- Operating conditions: dry, high S/N ratio, fast
- Communication: electronic
- Computing: digital, serial
- organization: Top-down

Material
Device
System

**Biological hardware:**
- High impedance
- Operating conditions: wet, low signal/noise ratio, slow
- Communication: Ionic, chemical
- Computing: analog, parallel
- organization: Bottom-up